\documentclass[final,5p,times,twocolumn]{elsarticle}
\newcommand{\pf}[2]{\frac{\partial ,1}{\partial ,2}}
\newcommand{\pfs}[2]{\frac{\partial^2 ,1}{\partial ,2 ^2}}
\newcommand{\pft}[2]{\frac{\partial^3 ,1}{\partial ,2 ^3}}
\newcommand{\pderiv}[2]{\frac{\partial ,1}{\partial ,2}}

\usepackage{lineno}
\usepackage{multirow}
\usepackage{tabu}
\usepackage{CJKutf8}
\usepackage[T1]{fontenc}
\usepackage[utf8]{inputenc}
\usepackage{graphics, amsmath}
\usepackage{graphicx,color}
\usepackage{epsfig}
\usepackage{amsmath}
\usepackage{amssymb}
\usepackage{setspace}
\usepackage{hyperref}
\usepackage{cleveref}
\usepackage{underscore}
\usepackage[dvipsnames]{xcolor}
\usepackage{multirow, makecell}
\usepackage{colortbl}
\usepackage{dashrule}
\usepackage{ehhline}
\usepackage{csquotes}
\usepackage{natbib}
\biboptions{sort&compress}

\usepackage{array}
\usepackage{booktabs}
\usepackage{caption}
\makeatletter
\usepackage{colortbl}
\usepackage{hhline}

\journal{Case Studies in Thermal Engineering}

\begin{document}
	
	\begin{frontmatter}
		
		\title{Thermal transport in thermoelectric materials of SnSSe and SnS$_2$: a non-equilibrium Monte-Carlo simulation of Boltzmann transport equation}
		\author[1]{Seyedeh Ameneh Bahadori}
		\author[1]{Zahra Shomali\corref{cor1}}
		\address[1]{Department of Physics, Basic Sciences Faculty, Tarbiat Modares University, Tehran, Iran}
		\cortext[cor1]{Corresponding author, Tel.: +98(21) 82883147.}
		\ead{shomali@modares.ac.ir}

		\begin{abstract}
			
			In the present work, thermal transport and the energy conversation in two thermoelectrically efficient candidates of Janus SnSSe and SnS$_2$ are investigated within the non-equilibrium Monte Carlo simulation of the phonon Boltzmann equation. The phonon analysis has been performed to determine the contributed phonon in heat transport. The results present that the dominant participating phonons are the ones from the longitudinal acoustic (LA) branch while the least belongs to the transverse acoustic (TA) modes. Both materials reached the very high maximum temperature in response to the implied wasted heat. This is attributed to the low presence of the critical TA phonons which are on one hand fast enough to leave the hotspot but on the other hand have frequencies commensurable to ZA phonons' frequency. Also, the temperature profile achieved during the heating and cooling of the materials is studied. It is obtained that the heat propagation through the SnS$_2$ is, at first, swifter, which results in a temperature gradient through the whole material which is less than that of the SnSSe. As the time passes, the heat transfer that is directly related to the material thermal conductivity, slows down. So, the behavior of the SnS$_2$ and SnSSe, in case of the heat propagation status, becomes similar. Moreover, the behaviour of the thermoelectric figure of merit (zT), the efficiency ($\eta$), and the generated voltage have been figured out. It is stated that the higher zT and $\eta$ do not guarantee a larger generated Seebeck voltage. This is true, while the generated Seebeck voltage is related to the temperature difference between the heated and the cold junction. Accordingly, how far the temperature of matter rises in response to the implied wasted heat is directly related to the obtained voltage. Mainly, it is presented that the maximum temperature that a material achieves, alongside the temperature gradient and the material property Seebeck coefficient, are essential in introducing thermoelectrically efficient materials with reasonable thermal to electrical energy conversion, useful for thermal engineering.
			
		\end{abstract}
		
	\end{frontmatter}
	

	\section{Introduction}
	
	Energy harvesting has been widely investigated since the early nineteenth. Nowadays, due to an urgent demand for the reduction of used nuclear, and fossil energy resources, the concept of renewable energy has been even more of interest \cite{DatD2020,Coulibaly2021,TDHong2023}. A complete replacement for the fossil fuels is not possible in the current situation and therefore, improving energy efficiency, resulting in reducing energy sources consumption is an alternative solution. Meanwhile, using the thermoelectric materials with the ability to convert the heat to electricity by the Seebeck effect and in reverse, driving the heat flow with an electric current called as the Peltier effect \cite{Zhang2015}, has been suggested as the way for promoting the energy performance. The thermoelectric material is the major part of a thermoelectric generator (TEG) \cite{Hussein2023,CGu2024}, which is a device having the potential to transform the waste heat into the electrical power. TEGs convert directly the temperature differences into the electric voltage \cite{T.M.Tritt2006}. As the voltage difference between the two nodes is linearly dependent on the amount of the temperature difference in between, the thermoelectric materials must have low thermal conductivity ($\kappa$), so, while the temperature on one side of the material has increased, the other side stays cold. This behaviour helps a generation of significant voltage. On the other hand, the thermoelectric efficiency of the thermoelectric materials relies on a parameter called the thermoelectric figure of merit (zT), which is determined using the formula zT=S$^{2}\sigma$T/K. In this relation, S, $\sigma$ and K, subsequently, represent the Seebeck coefficient, electrical conductance, and the thermal conductivity \cite{Snyder2017,Mishra2020}. The Seebeck coefficient measures the magnitude of an induced thermoelectric voltage in response to a temperature gradient across the material caused by the Seebeck effect. Further, the formula confirms the previously mentioned fact that decreasing the thermal conductivity, results in higher efficiency. The best thermoelectric performance measured until now, is reported in \cite{Hinterleitner2019}, which lies between 5 and 6. Additionally, the obtained voltage caused by the temperature gradient is defined as $\nabla$V=S$\nabla$T, with $\nabla$ T being the temperature gradient alongside the monolayer.
	
	In recent years, there have been many efforts to develop new thermoelectric materials for efficient power generation by increasing the zT. Specifically, the potential of low-dimensional materials has been widely discussed. For instance, 2D materials of stannane, silicene, phosphorene, bismuthene, MoS$_2$, MoSe$_2$, WSe$_2$, and SnSe have been proposed as practical thermoelectric materials due to their low lattice thermal conductivity, leading to the higher-value thermoelectric figure of merit, zT \cite{A.Patel2020,DLi2020}. Newlier, a more complex class of two dimensional materials, called Janus monolayers, as a new member of two-dimensional derivatives, have been given intense attention. These substances, with two layers of chalcogen atoms, built from two different elements, preserve the desirable properties of their parent binary structures but at the same time, use the additional degrees of freedom to further tuning the material characteristics. For instance, while dealing with the thermal transport, it has been obtained that the thermal conductivity of the 2D Janus materials of SnSSe, MoSSe, and ZrSSe is much lower than that of SnS$_2$, MoS$_2$, and ZrS$_2$ monolayers. This can be recognized by the traditional sense that a more complicated unit cell makes the phonon scattering increasing, and so inevitably the thermal conductivity reduces \cite{Guo2018}. The thermal conductivity reduction results in larger zT, which itself, in many literatures, is ascribed to the growth of the power generation. Accordingly, considering the low thermal conductivity and high zT together, the mechanically stable Janus monolayer compounds has been introduced as a promising candidate for thermoelectric applications \cite{Gupta2019,SBai2023}. Among Janus monolayers, SnSSe, derived from SnS$_2$ and SnSe$_2$, is a famous potential candidate in the thermoelectric field \cite{LiuG2020}.

	It is worthwhile to note that the stability status, such as the thermal stability, is a concern for the Janus materials. Specifically, among the Janus materials, SnSSe has a considerable maximum temperature at which it can stably exist. In fact, the analysis of the bond length fluctuations has demonstrated that the maximum temperatures up to which the Janus mono-layer SnSSe can stably exist, is as high as 825 K \cite{YLou2021}.  This is why this material has been suggested numerously as the possible candidate. Also, the dynamic stability for SnSSe has been verified in \cite{PWang2020}. Mainly, various ternary Janus metal dichalcogenides like {Mo,Zr,Pt}-SSe have been introduced as the possible candidates for the thermoelectric applications. However, among the aforementioned nominees, SnSSe, has been shown recently to be mechanically stable \cite{SDGu2019,RGupta2020}. Hence, alongside, the monolayer SnS$_2$ with a much lower lattice thermal conductivity in comparison with that of the other conventional transition metal dichalcogenides, SnSSe has been also proposed as the efficient thermoelectric material. Due to the importance of the thermoelectricity and also the significant role of SnSSe and SnS$_2$ as the thermoelectric material candidates, their comparative study, focusing on how the heat transport occurs in, becomes consequential. 
	
	In more details, performing phonon analysis which reveals the dominant heat carriers’ type, alongside tracing the temperature and heat flux behavior, leads to valuable knowledge about the heat distribution state, and the temperature value and arrangement, inside the favorite materials. The precise calculation of the temperature inside the thermoelectric material is an indispensable concept, while finding the accurate values of the critical thermoelectric parameters such as output voltage and the figure of merit, directly depends on the temperature. On the other hand, the low thermal conductivity characteristic of the thermoelectric materials makes the heat spreading within the matter pretty difficult and the TEG encounters practical limitations. Consequently, uncovering the temperature behavior inside the thermoelectric material, for instance, finding out the maximum reached temperature and the temperature gradient, under the influence of the waste thermal energy is crucial for thermoelectric material solution.
	
	 On the other hand, the thermoelectric figure of merit zT has been widely used as an indicator of the thermoelectric material status, such that, the materials with high zT, like the newly proposed Janus monolayers, are recommended as the thermoelectrically efficient materials. In other words, for decades, there were attempts for searching higher zT values and accordingly the zT value has been improved over time. It was for the first time in 2000 that the T exceeded 1 for the first time due to the finding of new nanostructures with low thermal conductivity. In spite of great attention to the zT value, there have been doubts about the direct relation of the zT and the efficiency. Further, the figure of merit, zT, has been extensively used as a simple estimator of the conversion efficiency of a thermoelectric generator. The calculated efficiency is proportional to the average zT over the whole range of the temperature and consequently can not predict the real situation. Basically, as the material properties are kept constant or slowly varying with the temperature, one can say that a higher zT guarantees a larger maximum thermoelectric conversion efficiency. However, the material properties can be strongly temperature dependent and so results in inaccurate zT efficiency predictions \cite{HSKim2015,BRyu2020,BRyu2023}. This finding leads one to the point that the zT value is not the only important parameter in introducing new appropriate thermoelectric materials and one should excavate for uncovering the more involved requirements. The present paper, using the fundamental thermal parameters such as the phononic dispersion curves and the temperature/frequency dependent scattering rates, tries to investigate the criteria, which are alongside the zT, responsible for more efficient thermoelectrically situation, including the higher efficiency and larger voltage.
	
	The temperature profile, phonon analysis, and the transient heat transport inside a thermoelectric material can be obtained using the macroscopic or atomistic models \cite{Ghazanfarian2009,Shomali2012}. Here, the atomistic method of non-equilibrium Monte Carlo simulation of the Phonon Boltzmann Equation (PBE) is utilized. As the PBE takes into account the atomic characteristics such as the phonon dispersion relation and the relaxation times, representative of the different scatterings occurring, the temperature distribution can be thoroughly found through the material. In this research, two materials of Janus SnSSe and SnS$_2$ with respectively low thermal conductivity of 3.73 and 0.61 Wm$^{-1}$K$^{-1}$ at room temperature, have been studied. For the studied materials, the dominant contributed phonons in thermal transport, likewise the most two-dimensional materials, belong to the acoustic branches. In the present work, first, the contribution of each acoustic branch in transient heat transfer is studied. Particularly, the phonon analysis, to uncover the status of the heat carriers inside the both materials, has been carried out. Then, the temperature profile inside the two thermoelectric materials of SnS$_{2}$ and Janus SnSSe has been obtained. At the next step searching for the status of the maximum temperature reached in two studied monolayers, the SnS$_2$ two-dimensional monolayer has been found to present very much little higher maximum temperature in response to the same waste heat applied. This fact, in beginning, results in an almost larger temperature gradient between the hot and cool sides. For a subsequent stage, the figure of merit, zT, and the lattice thermal conductivity have been calculated for both monolayer two-dimensional materials. The results for zT verify the previous studies, confirming that the parameter zT for SnSSe is always almost three times larger than that of the SnS$_{2}$ at every temperature. Moreover, the temperature gradient alongside the x-direction, and the thermoelectric efficiency of the monolayers, beside the generated voltage difference is investigated. Correspondingly the obtained results show that while the SnSSe monolayer has much larger zT, but in practice, the energy conversion status is not very far from that of the SnS$_2$ thermoelectric material. In this paper, in Sec. \ref{Sec.2}, the studied geometry and the relevant boundary conditions are discussed. Sections. \ref{Sec.3} and \ref{Sec.4}, respectively, deal with introducing the utilized mathematical modelling and the numerical method. At last, the results are given in Sec. \ref{Sec.5} and, the paper is concluded in Sec. \ref{Sec.6}.
	
	\section{Geometry and Boundary conditions}
	\label{Sec.2}
	Here, two monolayers of SnS$_2$ and Janus SnSSe which are suggested efficient thermoelectric materials have been investigated. The optimized structure of the Janus SnSSe and SnS$_2$ monolayers have a similar form. Nevertheless, in SnSSe, the layer of Sn atoms is placed between the S and Se with a reflection symmetry broken along them. The studied monolayer materials have dimensions of width, L$_x$=100 nm, and length, L$_y$=100 nm. Also, the attributed thickness for SnSSe and SnS$_2$ materials is, respectively, 0.678 and 0.655 nm. The materials are under the influence of the implied waste heat. In more detail, the homogeneous heat flux of Q=1$\times$10$^{12}$ W/m$^{3}$ is applied along the left boundary of the channel where the position alongside the x-direction is between zero and 10 nm. All boundaries excluding the bottom boundary, are considered adiabatic with the temperature being equal to the ambient temperature. The bottom boundary makes the investigated system exchange the energy with the environment \cite{Shomali2017,2Shomali2017}. It must be pointed out that the initial temperature is kept at T= 299 K. 
	In addition, the low-dimensional channels in our simulation are rectangular and considered flat edges with specular reflection. With this type of reflection from the flat edges, the phonon velocity component along the substance will not change and so there will be no resistance.
	
	\section{Mathematical Modeling}
	\label{Sec.3}
	Here, as the heat transport at nanoscale is dealt, one should imply the non-Fourier methods \cite{Shomali20152,Shomali2016,Liu2020,Shomali2021,Shomali2022,Shomali2024}. In other words, the thermal transport through the nano materials is expressed via the behavior of the phonons. Hence, in this study, the phonon Boltzmann transport equation is solved to determine the phonon distribution function, f$_b$$(\mathbf{r},\mathbf{q},t)$, with parameters $\mathbf{r}$, $\mathbf{q }$, and t being respectively the location, the wave vector, and time for each phonon branch of b \cite{Shomali2019,Sattler2022}. Also, the phonon Boltzmann transport equation (PBTE) is written as
	
	\begin{equation}
		\frac{\partial \textrm{f}_{b}(\mathbf{r},\mathbf{q},t)}{\partial t}+\textrm{v}_{b,\mathbf{q}} \ . \ \nabla_{r}\textrm{f}_{b}(\mathbf{r},\mathbf{q},t)=\frac{\partial \textrm{f}_b (\mathbf{r},\mathbf{q},t)}{\partial t}\mid_{scat}.
	\end{equation}
	
	In the above relation, the phonons group velocity and the angular frequency of the branch $b$ with the wave vector $\mathbf{q}$ are presented as v$_{b,\mathbf{q}}$, and $\omega$$_{b,\mathbf{q}}$. The frequency $\omega_{b,\mathbf{q}}$ relies on the magnitude and the direction of the wave vector \cite{THLiu2015}. In the current study, the nonequilibrium Monte Carlo (NMC) method is employed to simulate the PBTE. The SnSSe, and SnS$_2$, likewise the other low-dimensional materials, have six phononic acoustic and optical branches while only the acoustic phonons are the one which are important in thermal transport. Moreover, the acoustic phonon dispersion curves are almost isotropic and can be fitted to the quadratic equations \cite{Ge2016}. In consequence, the quadratic polynomial of the form $\omega$$_b$=c$_{b}$k$^{2}$+v$_{b}$k is tried out for the two studied materials of SnSSe and SnS$_2$. The obtained fitting coefficients are reported in the Table. \ref{Tab1-tab1}.	
	
	In the path for simulation, the frequency interval being started from the zero to the maximum value of the frequency for each acoustic dispersion curve $\omega_{max}$, will be divided by N$_{\rm int}$=1000 steps. Here, the discretized frequency range, $\Delta\omega_{i}$, will be considered as 4.039$\times$10$^{12}$, and 4.946$\times$10$^{12}$ for SnSSe and SnS$_2$ monolayers \cite{Shomali2018,2Shomali2017}.
	
	\begin{table}[htbp]
		\caption{The values of the coefficients for the fitted quadratic formula, $\omega$$_b$=c$_b$k$^{2}$+v$_b$k, for SnSSe and SnS$_2$ low-dimensional materials.}
		\label{Tab1-tab1}
		\hspace{-0.8cm}
		\vspace{-0.5cm}
		\begin{center}
			\begin{small}
				\begin{tabular}{cccc}
					\hline
					2D Material    &  SnSSe    &
					SnS$_2$    \\ \hline \hline
					c$_{LA}$(m$^2$/s)  & -3.904$\times$10$^{-8}$ & -4.569$\times$10$^{-8}$ \\ \hline
					c$_{TA}$(m$^2$/s) & -1.928$\times$10$^{-8}$ & -2.408$\times$10$^{-8}$ \\ \hline
					c$_{ZA}$(m$^2$/s) & -4.180$\times$10$^{-9}$ & -5.909$\times$10$^{-9}$ \\ \hline
					v$_{LA}$(m/s) & 3992 & 4788  \\ \hline
					v$_{TA}$(m/s) & 2521 &  3052 \\ \hline
					v$_{ZA}$(m/s) & 263 & 139 \\ \hline
				\end{tabular}
			\end{small}
		\end{center}
	\end{table}
	
	During the simulation, when the step for scattering is performed, the phonons may experience collision with other phonons or scattering from the boundaries \cite{Shomali2018,2Shomali2017}. In this paper, the boundaries are taken to be flat and so they dictate a specular reflection. When the temperature range during the simulation is kept from 300 K to 700 K, the most essential mechanism for a low-dimensional suspended system is the phonon–phonon scattering. Here, two mechanisms of three phonon Umklapp/normal scattering and the boundary scattering, are the ones which are taken into account during the simulation. In this place, to treat the scattering, an approximation due to the nearly high temperature treatment is applied \cite{Mazumder2001,Cepellotti2017}. According to this appropinquation, the Umklapp phonon-phonon scattering rate will be estimated using the standard general approximation for dielectric crystals formula, $\tau^{-1}_{b,U}(\omega)=\frac{\hbar \gamma ^{2}_{b}}{\bar{M} \Theta_{b} v^{2}_{s,b}} \omega^{2} T e^{-\Theta_b/3T}$ \cite{THLiu2015}. In the mentioned expression, the parameters v$_{s,b}$, $\bar{M}$, and $\Theta$, are, subsequently, the branch b sound velocity, the average atomic mass, and the Debye temperature. It is worth noting that the starting phrase of the equation is responsible for the standard Umklapp interaction strength, while the exponential part describes the contribution of redistribution by the N processes. The value of $\Theta$, $\bar{M}$ and Gr\"{u}neissen parameters, $\gamma$, used throughout the simulation, are declared in Table. \ref{Tab2-BP}.
	
	\begin{table}[htbp]
		\caption{The thermal characteristics utilized for calculation of the relaxation times. \newline}
		\label{Tab2-BP}
		\vspace{-0.5cm}
		\centering
		\begin{small}
			\hspace*{-0.4cm}
			\begin{tabular}{cccc}
				\hline
				2D Material &  SnSSe    &
				SnS$_2$   \\ \hline \hline
				$\Theta$  & 190  & 233  \\ \hline
				$\bar{M}$(e$^{-27}$ kg) & 127.2 & 101.13 \\ \hline
				$\gamma$ &  1.6 & 1.48 \\ \hline
			\end{tabular}
		\end{small}
	\end{table}
	
	\section{Numerical method}
	\label{Sec.4}
	The mesh independency test, performed for the simulation area, has shown that a uniform mesh with the size of $100\times $100 in the XY-plane, is the appropriate choice for obtaining the mesh-independent plots. On the other side, to reach the adequate time step, first the frequency range of all three acoustic branches is divided into 1000 intervals. Then, the attributed velocity for each frequency differential is calculated. In the next step, dividing the mesh size by the velocity of each step, the phonon travel time is obtained. Finally, going through the whole frequency range, the minimum calculated travel time will be selected as the time step with the condition of holding the time independence. More particularly, the time steps of 2.514$\times$10$^{-12}$, and 2.1$\times$10$^{-12}$ s, are acquired for SnSSe, and SnS$_2$. 
	Furthermore, the scattering rates for phonons with all 1000 possible frequencies in each branch are also searched out. Accordingly, the scattering rates will be determined by reversing the phonon relaxation times. At last, it is worth mentioning that the wasted heat source area is applied from the left side of the materials into a 10$\times$100$\times$0.65 nm$^{3}$ cube of simulation area. The implied wasted heat until when it reaches the desired value, will be made by releasing the phonons into the region of heating. 	
	\section{Results and Discussions}
	\label{Sec.5}
\begin{figure}
		\vspace{-0.88cm}
		\centering
		\includegraphics[width=\columnwidth]{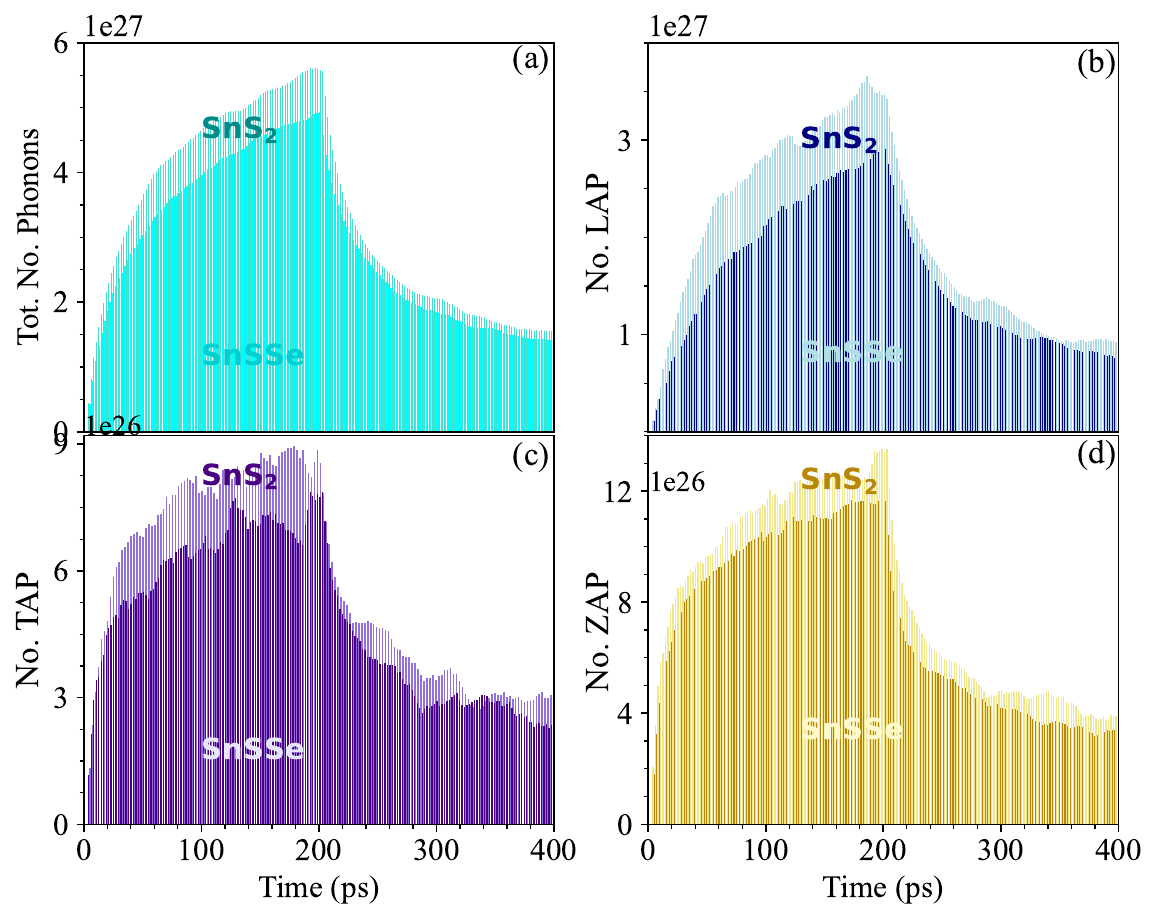}
		\caption{\label{contribution-of-branches-1} The number of phonons per volume behavior, for two materials of SNSSe and SnS$_2$ versus the time for (a) all branches together, and (b) LA, (c) TA, and (d) ZA branches.}
	\end{figure}

\begin{figure}
		\vspace{-1.0cm}
		\centering
		\includegraphics[width=\columnwidth]{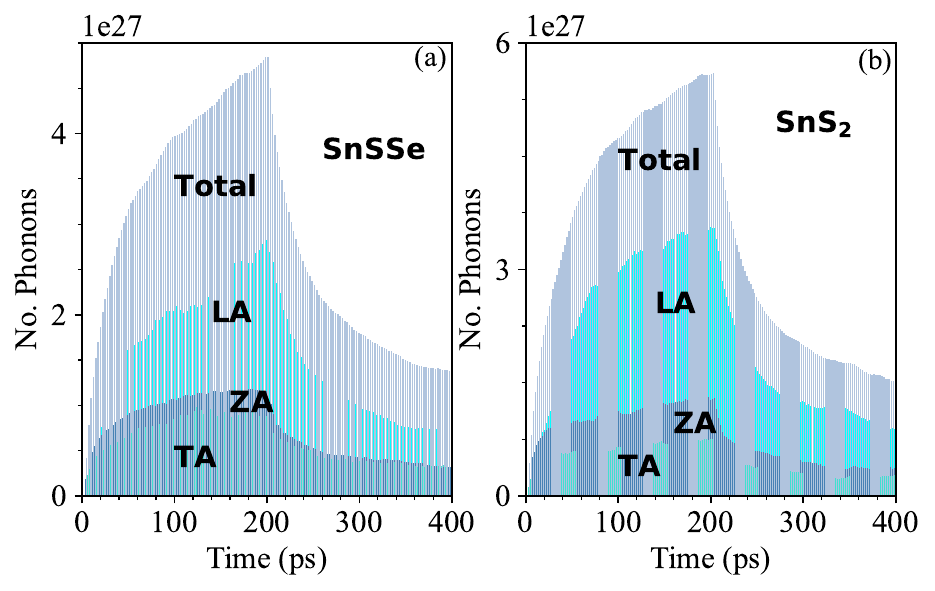}
		\caption{\label{contribution-of-branches-2} The time plot bar graph for participation percentage of LA, TA, and ZA phonons when the monolayer material is (a) SnSSe, and (b) SnS$_2$.}
	\end{figure}

	In the present study, the non-equilibrium Monte Carlo simulation of the phonon Boltzmann transport equation has been implemented for obtaining the transient heat transport circumstances inside the monolayers of SnSSe and SnS$_2$. The monolayers are heated through the first 200 ps of the simulation, and then the heat generation zone is turned off. The simulation primarily deals with calculating the contribution percentage of the phonons in different branches. The results are presented in Table. \ref{contribution-of-branches}, and Figs. \ref{contribution-of-branches-1}, and \ref{contribution-of-branches-2}. As it is shown in Fig. \ref{contribution-of-branches-1}, the total number of the phonons taking part in the heat transport through the SnSSe monolayer, is always slightly more than that of the SnS$_2$ during the heating and cooling.
	
\begin{figure}
		\vspace{0cm}
		\centering
		\includegraphics[width=\columnwidth]{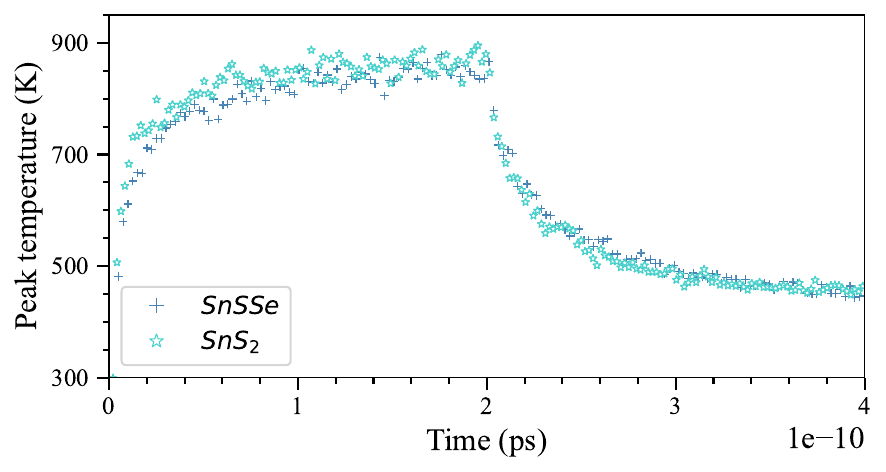}
		\caption{\label{peaktemp} The maximum temperature that the SnSSe and SnS$_2$ monolayers achieve during the heating and the cooling versus the time.}
	\end{figure}

	Moreover, as the table. \ref{contribution-of-branches} suggests, although at the first few picoseconds, the dominant phonons taking part in heat transport in both monolayers are the slow-low frequency ZA phonons, however they promptly switch to the fast-high frequency LA phonons. It is also worthwhile to mention that the TA phonons play the least role in thermal transport through the SnSSe and SnS$_2$. The transverse acoustic phonons are the ones which are involved in thermal management solutions. This is since that the TA phonons, on one hand, are fast enough to free themselves out of the hotspot, and on the other hand, possess frequencies comparable to that of the ZA phonons and less than the LA mode frequencies. The lower frequency results in lesser carrying energy and consequently, lower temperature. Therefore, the lack of TA phonons makes the maximum temperature increase. Figure. \ref{contribution-of-branches-2} presents that the most dominant phonons participating in heat transport for both studied materials at all times are, subsequently, LA, ZA, and TA phonons. The contribution of the phonons in TA frequency are moderately larger for SnSSe monolayer. In more detail, as it is manifested in the Table. \ref{contribution-of-branches}, the contribution of the TA phonons for SnSSe and SnS$_2$, respectively, reduces down to 17$\%$ and 13$\%$ for the time period of 150-250 ps. In consequence of how different kinds of phonons are distributed, both monolayers of SnSSe and SnS$_2$ are expected to reach a high peak temperature while that of the SnSSe is a bit smaller. This anticipation is confirmed in Fig. \ref{peaktemp}. More particularly, the maximum temperature that the materials reach at every time, is exhibited in Fig. \ref{peaktemp}. The high peak temperature rise together with the large value steady state temperature is noticeable. It should be mentioned that the high maximum temperature is the positive point toward the usage of the material as a thermoelectric one, while making it an unsuitable one for a MOSFET channel. In other words, the large maximum temperature generating the higher voltage difference, beside the low thermal conductivity of the monolayers SnSSe and SnS$2$, makes the materials, the efficient ones for thermoelectric utilization. 
	
	\begin{table*}[htbp]
		\vspace{-0.35cm}
		\begin{center}
			\begin{small}
				\begin{tabular}{|c|c|c|c|c|c|c|}
					\hline
					\multirow{2}*{t (ps)}&\multicolumn{3}{c} {\cellcolor{Aquamarine!10}(a) SnSSe} &	\multicolumn{3}{ |c| } {\cellcolor{Aquamarine!10}(b) SnS$_2$} \\	
					\cline{2-6} \	
					& {\cellcolor{Aquamarine!20}LA (\%)}& {\cellcolor{Aquamarine!25}TA (\%)}&{\cellcolor{Aquamarine!30}ZA (\%) }& {\cellcolor{Aquamarine!20}LA (\%)} & {\cellcolor{Aquamarine!25}TA (\%)}&{\cellcolor{Aquamarine!30}ZA (\%)}\\
					
					\hline 
					\cellcolor{purple!5}10 &  {\cellcolor{cyan!5}29} &  {\cellcolor{cyan!15}31}& {\cellcolor{cyan!25}40} & {\cellcolor{RoyalBlue!7}22} & {\cellcolor{RoyalBlue!15}29}& {\cellcolor{RoyalBlue!25}49}\\
					\hline
					\cellcolor{purple!10}50 &  {\cellcolor{cyan!25}49} & {\cellcolor{cyan!5}22} & {\cellcolor{cyan!15}29} & {\cellcolor{RoyalBlue!25}52} &  {\cellcolor{RoyalBlue!7}18} & {\cellcolor{RoyalBlue!15}30}\\
					\hline
					\cellcolor{purple!15}100 &  {\cellcolor{cyan!25}58} &{\cellcolor{cyan!5}18} & {\cellcolor{cyan!15}24} &  {\cellcolor{RoyalBlue!25}61} & {\cellcolor{RoyalBlue!7}14}& {\cellcolor{RoyalBlue!15}25}\\
					\hline
					\cellcolor{purple!20}150 &  {\cellcolor{cyan!25}59} & {\cellcolor{cyan!5}17}& {\cellcolor{cyan!15}24} &  {\cellcolor{RoyalBlue!25}63} &{\cellcolor{RoyalBlue!7}13} & {\cellcolor{RoyalBlue!15}24}\\
					\hline
					\cellcolor{purple!25}200 &  {\cellcolor{cyan!25}60} & {\cellcolor{cyan!5}17} & {\cellcolor{cyan!15}23} & {\cellcolor{RoyalBlue!25}65} & {\cellcolor{RoyalBlue!7}13}& {\cellcolor{RoyalBlue!15}22} \\
					\hline
					\cellcolor{purple!30}250 &  {\cellcolor{cyan!25}62} & {\cellcolor{cyan!5}17}& {\cellcolor{cyan!15}25} & 	 {\cellcolor{RoyalBlue!25}66} &{\cellcolor{RoyalBlue!7}13} & {\cellcolor{RoyalBlue!15}21}\\
					\hline
					\cellcolor{purple!35}300 &  {\cellcolor{cyan!25}59} & {\cellcolor{cyan!5}18}& {\cellcolor{cyan!15}23} &  {\cellcolor{RoyalBlue!25}62} & {\cellcolor{RoyalBlue!7}15}& {\cellcolor{RoyalBlue!15}23} \\
					\hline
					\cellcolor{purple!40}350 &  {\cellcolor{cyan!25}58} & {\cellcolor{cyan!5}19}& {\cellcolor{cyan!15}23} &  {\cellcolor{RoyalBlue!25}61} & {\cellcolor{RoyalBlue!7}15}& {\cellcolor{RoyalBlue!15}24}\\
					\hline
					\cellcolor{purple!45}400 &  {\cellcolor{cyan!25}57} & {\cellcolor{cyan!5}19}& {\cellcolor{cyan!15}24} &  {\cellcolor{RoyalBlue!25}59}  &{\cellcolor{RoyalBlue!7}16} & {\cellcolor{RoyalBlue!15}25} \\
					\hline 
				\end{tabular}
			\end{small}
		\end{center}
		\vspace{-0.5cm}
		\hspace{-1.7cm}	\caption{\label{contribution-of-branches}The contribution of LA, TA, and ZA during the heat transport in materials of (a) SnSSe, and (b) SnS$_2$. The set of the colors (light to dark) presents the branches with the least to a maximum participation.}
	\end{table*}
	
\begin{figure*}
		\vspace{-1.61cm}
		\centering
		\includegraphics[width=2\columnwidth]{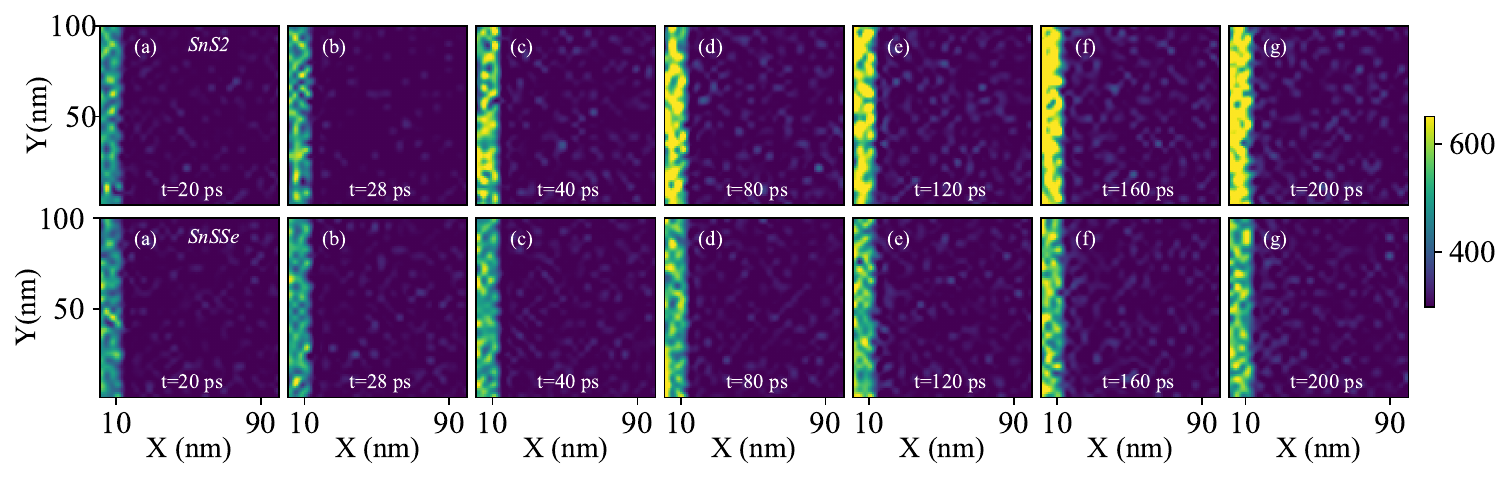}
		\caption{\label{Tprofile-heating} The hotspot behaviour at XY plane during the heating when t= (a) 2, (b) 10, (c) 50, (d) 100, (e) 150, (f) 170, and (g) 200 \textbf{ps} for SnSSe and SnS$_2$ monolayers appearing, respectively, in the upper and lower plots.}
	\end{figure*}

	Furthermore, the behavior of the temperature profile during the heating and cooling are also investigated. As previously mentioned, the waste heat is implied from the left side. Despite the temperature distribution at the first few picoseconds where both monolayers have nearly the same profile due to the higher amount of the TA phonons relative to the LA ones, as time passes, the overall temperature through the SnS$_2$ becomes partially larger. The hot spot formation during the first 200 ps is shown in Fig. \ref{Tprofile-heating}. It is evident that the hot spots are formed immediately after the external heat is applied. As the Figs. \ref{Tprofile-heating} (a-g). confirm, at first, the heat propagation in SnS$_2$ is more pronounced and also swifter. For instance, by paying attention to upper and lower Figs. \ref{Tprofile-heating} (d) and (f), one can compare the temperature profiles for SnS$_2$ and SnSSe when the time is t=80 and 120 ps. It is evident that 80 ps after turning on the heating zone, for the material SnS$_2$, the heat propagation has influenced nearly all the sheet while the effect is more noticeable for X=10-25 nm. This is while the lower Figs. \ref{Tprofile-heating} (d) and (f) establish that the heat flow alongside the SnSSe material is slower. In this case the gradient between the temperature of the hot and cold zone becomes more prominent and this is why one expects that the SnSSe should be a more efficient thermoelectric material. This behavior is also attributed to the lower thermal conductivity of the SnSSe. On the contrary, one should be attentive that the lower thermal conductivity is not only the term which distinguishes the heat propagation. As described before, the phonon analysis should also be performed to see the percentage of the contribution of each type of the phonon. This is important while it specifies the temperature that the hot region attains during the effectuation of the turned on applied external heat. It is indisputable that increasing the maximum temperature at hot region concurrently with the remarkable temperature gradient inside the material due to the slowness of the heat propagation, are two significant factors that determine the efficiency of the thermoelectric candidate. At the same time, as the temperature increases, the thermal conductivity of the SnS$_2$ decreases much more than the one for SnSSe. So, as the time passes and consequently the temperature augments, the heat propagation through the SnS$_2$ becomes slower and the conditions become more similar to what exist in SnSSe except that it has reached higher temperature in response to the implied heat.
	
\begin{figure*}
		\centering
		\includegraphics[width=2\columnwidth]{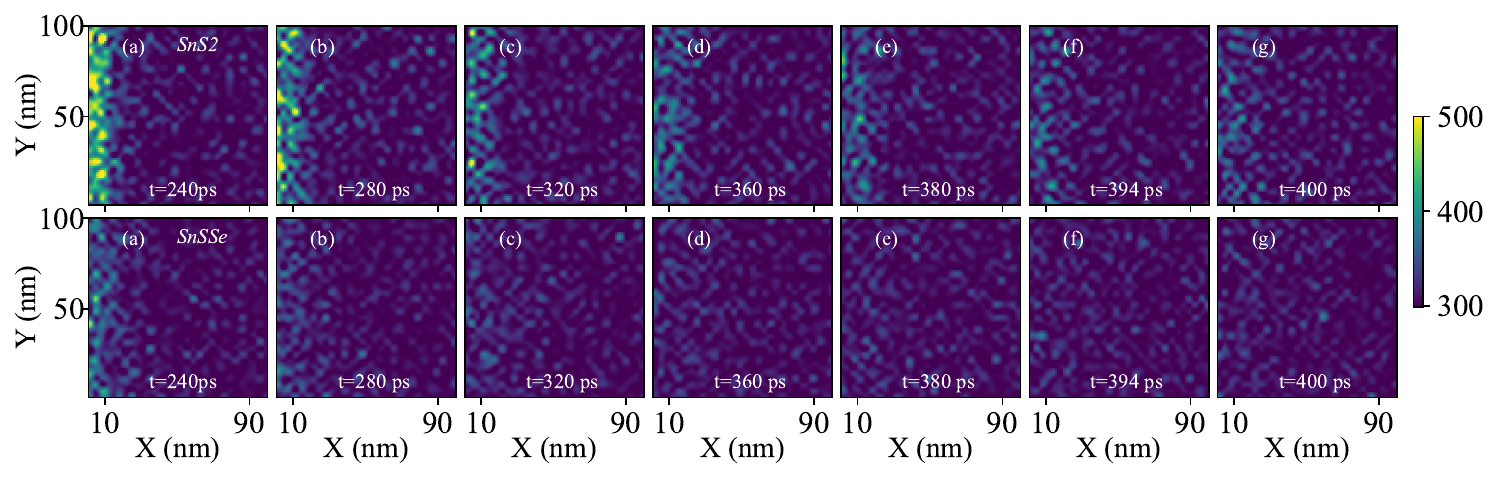}
		\caption{\label{Tprofile-cooling} The temperature distribution at XY plane while the monolayers are being cooled at t= (a) 210, (b) 250, (c) 300, (d) 330, (e) 360, (f) 380, (g) 400 \textbf{ps} for SnSSe and SnS$_2$ materials, subsequently, presented in upper and lower graphs.}
	\end{figure*}

	In the same way, the state of the temperature distribution after turning off the heat-implemented region is investigated. The bottom boundary is also considered open, exchanging energy with the surroundings. The results are manifested in Figs. \ref{Tprofile-cooling} (a-g). As previously stated, the dominant phonons taking part in the heat transport are still LA and TA phonons which are fast enough to leave the hotspot in lower time. More specifically, although the thermal conductivity of the SnSSe is less than that of the SnS$_2$, due to the lower hot region temperature, the form of the hotspot is ruined and the material reaches the steady state in lesser time and temperature. This is true while it takes much time for SnS$_2$ to achieve the steady state. This can be attributed to the conformation of the hotspot with the higher temperature and also decreasing the thermal conductivity with increasing the temperature for the SnS$_2$ monolayer. 
	
	
\begin{figure}
		\centering
\hspace{-0.5cm}
		\includegraphics[width=1.0\columnwidth]{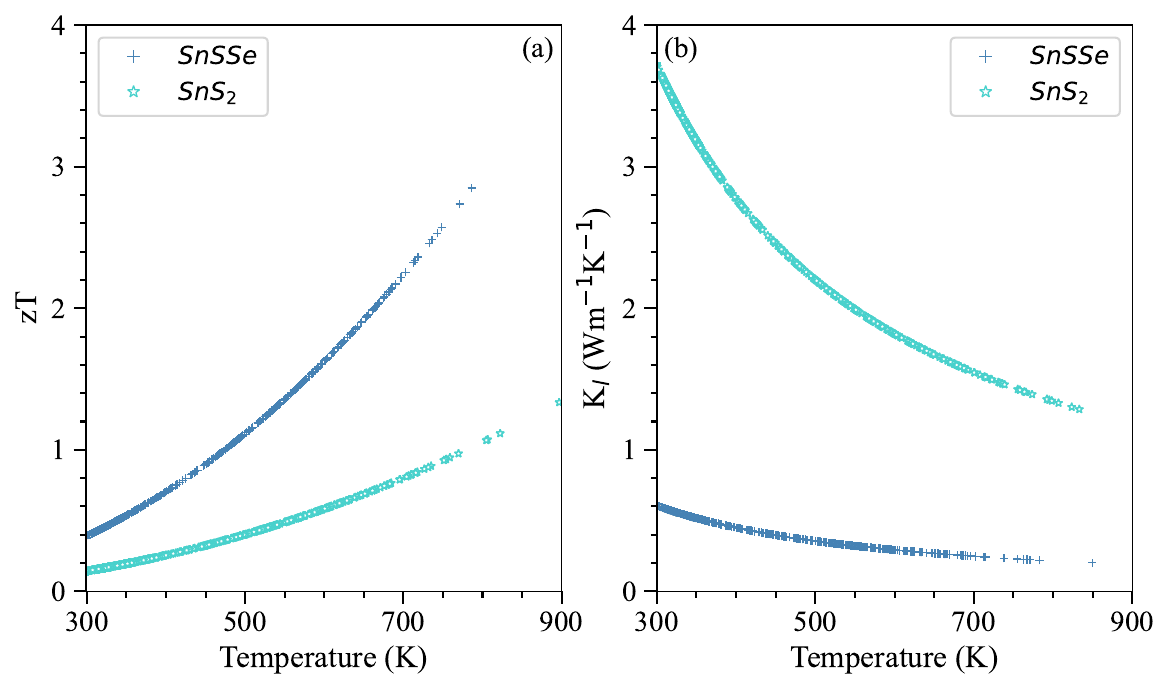}
		\caption{\label{kl-zt} The obtained (a) figure of merit (zT) and (b) the lattice thermal conductivity versus the temperature for both two materials of SnS$_2$ and SnSSe.}.
	\end{figure}

	Since the relationship between the thermoelectric figure of merit, zT, and also the lattice thermal conductivity and temperature becomes important for thermoelectric efficiency analysis, here, these two parameters are calculated. As the Figs. \ref{kl-zt} (a) and (b) present, the difference between the obtained zTs for SnSSe and SnS$_2$ always increases with temperature such that for T=800 K, the SnSSe figure of merit becomes almost three times larger than that of the one for SnS$_2$. The finding are consistent with what previously obtained in \cite{Nautiyal2022}. It is also worth pointing out that the quadratic formula can be fitted to the zT plots. The a,b, and c coefficients of the formula "a+bT+cT$^2$" are given in Table. \ref{zT}. Also, on the other side, the lattice thermal conductivity of the two monolayer materials has been calculated. It is obtained that the difference between the thermal conductivity of the SnSSe and SnS$_2$ is very high at room temperature. Although this discrepancy becomes lower as time goes by, it is shown that the K$_L$ for SnSSe remains during the simulation around six times larger than the SnS$_2$ lattice thermal conductivity. As already indicated, the thermoelectric figure of merit is calculated via the formula zT=S$^2$$\sigma$/(K$_L$+K$_e$). Where, S$^2$$\sigma$, K$_L$, and $K_e$ are, subsequently, the power factor, the thermal and electrical conductivity. As presented in \cite{Nautiyal2022}, the power factor for SnS2 is nearly 1.7 times larger than that of the SnSSe for hole doped and also electron thermal conductivity, for both SnSSe and SnS$_2$ is almost of the same order. Consequently, the behavior of the zT obeys the inverse attitude of the lattice thermal conductivity \cite{Nautiyal2022}. On the other hand, the lattice thermal conductivity is calculated via K$_L$=$\frac{1}{3}$C$_v$ v$_{avg}$$l$. In the expression, C$_v$ and v$_{avg}$ are the specific heat and the average sound velocity. The parameter $l$ is the phonon mean free path. For two studied materials of SnSSe and SnS$_2$, as Table. \ref{Tab1-tab1} suggests the average sound velocity for SnSSe is lower. In addition, using the standard general approximation formula for the scattering rate, the SnS$_2$ monolayers are obtained to have a notably higher phonon lifetime than the Janus monolayer. For instance, at T=300 K, the phonon lifetime for SnS$_2$ is almost two times that of the SnSSe. Thereby, the lower phonon lifetime beside the lower average sound velocity results in lower lattice thermal conductivity for Janus monolayers and explain why these materials present ultralow lattice thermal conductivity. As already noted, the lower lattice thermal conductivity means higher zT and that is why SnSSe is proposed as the possible efficient thermoelectric material. Accordingly, the thermal conductivity value in cooperation with the type of the phonons participating in the heat transport, are two important parameters that determine the peak temperature rise. 
	
	\begin{table}[htbp]
		\caption{The coefficients for the zT fitted quadratic formula, a+bT+cT$^{2}$, for the materials SnSSe and SnS$_2$.}
		\label{zT}
		\hspace{-0.8cm}
		\vspace{-0.5cm}
		\begin{center}
			\begin{small}
				\begin{tabular}{cccc}
					\hline
					2D Material    &  SnSSe    &
					SnS$_2$    \\ \hline \hline
					a  & 3.83$\times$10$^{-1}$ & 1.42$\times$10$^{-1}$ \\ \hline
					b & 1.72$\times$10$^{-3}$ & 6.02$\times$10$^{-4}$ \\ \hline
					c& 2.3$\times$10$^{-6}$ & 8.28$\times$10$^{-7}$ \\ \hline
					
				\end{tabular}
			\end{small}
		\end{center}
	\end{table}
	
\begin{figure}
		\centering
		\includegraphics[width=1.0\columnwidth]{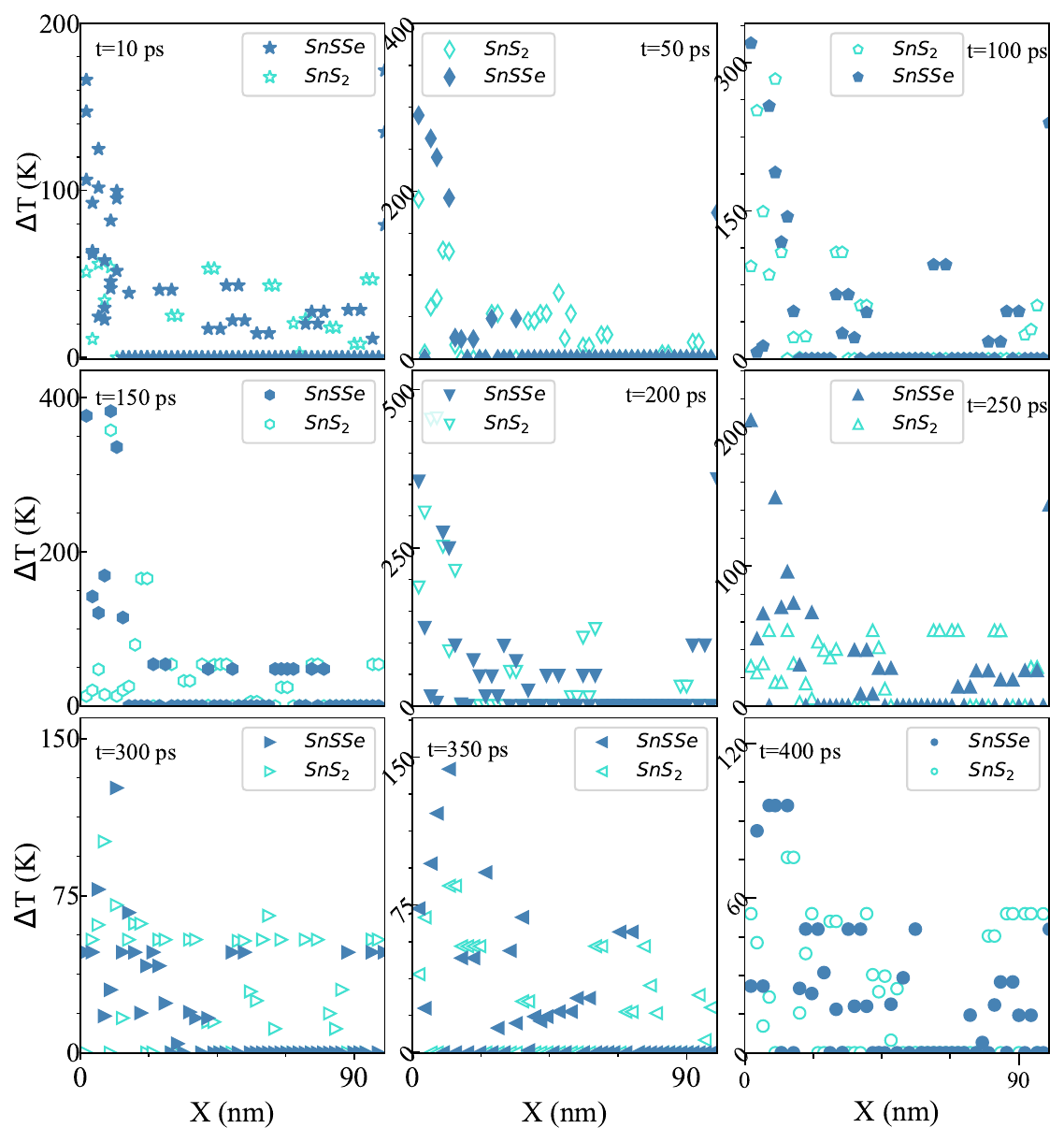}
		\caption{\label{vol} The behaviour of the temperature difference between the two neighboring nodes alongside the x-direction when x=0-100 nm and y=50 nm for the SnS$_2$ and SnSSe.}
	\end{figure}

	Taking into account the behavior of the zT and K$_L$, it is important to note that the temperature gradient induced in the material, is also responsible for thermoelectric efficiency. Indeed the maximum thermoelectric efficiency is defined through the formula $\eta_{max}=\frac{\Delta T}{T_H}\times\frac{\sqrt{1+ZT_{avg}}-1}{\sqrt{1+ZT_{avg}}+\frac{T_c}{T_H}}$. Where, T$_H$ and T$_c$ present, respectively, the hot and cold temperature, and T$_{avg}$ and $\Delta$T are defined, subsequently, as (T$_H$+T$_C$)/2, and T$_H$-T$_C$. Correspondingly, in the following step, the status of the temperature gradient inside the material has been investigated. As indicated earlier, the width of the material has been discretized into 100 nodes. The Fig. \ref{vol} presents the difference between the temperature of the two neighboring nodes along the X-direction within X=0-100 nm and in the middle of the monolayers at Y equal to 50 nm. As the Fig. \ref{vol} implies, in case of considering the node-by-node temperature difference, the gradient has an almost slighter higher value for the SnSSe material. During the simulation, say from 50-200 ps, when the time progresses and consequently the temperature increases, mostly in the left side where the heat is applied, the thermal conductivity for SnS$_2$ decreases more. This is while simultaneously the heat flow in the righter side with lower temperature is faster reaching the right boundary. This causes the temperature gradient in SnS$_2$, also increase. Although our focus is on the thermoelectric behavior in response to the applied waste heat, but the plots for times larger than 200 ps, when the heating source is turned off, are also presented to obtain a complete view about the heat transport in the studied monolayers. In fact, when the heat source is switched off, the temperature through the materials decreases and consequently the thermal conductivity is reduced, and so the temperature gradient becomes less until it reaches a steady state. This figure is presented to confirm the claim that the temperature gradient through the SnSSe monolayer should be higher in the first moments. 
	
\begin{figure}
		\hspace{-0.28cm}
		\centering
		\includegraphics[width=\columnwidth]{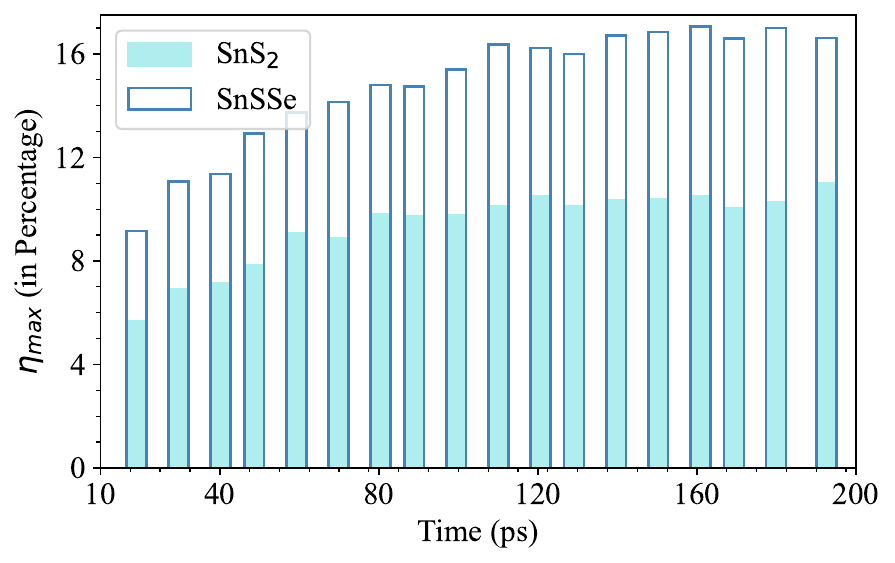}
		\caption{\label{efficiency} The maximum thermoelectric efficiency, $\eta_{max}$, presented in percentage, $\%$, versus the time for SnSSe and SnS$_2$.}
	\end{figure}

	Furthermore, the maximum thermoelectric efficiency has been calculated. The formula is dependent on the cold and hot side's temperature and also the difference between them, as well as the zT for the average temperature, (T$_H$+T$_C$)/2. The plot bar showing the average maximum efficiency for each time within the heating appears in Fig. \ref{efficiency}. The results present the existence of the efficiency which is always more prominent for SnSSe due to the higher averaged figure of merit. 
	
\begin{figure}
		\hspace{-0.28cm}
		\centering
		\includegraphics[width=1.05\columnwidth]{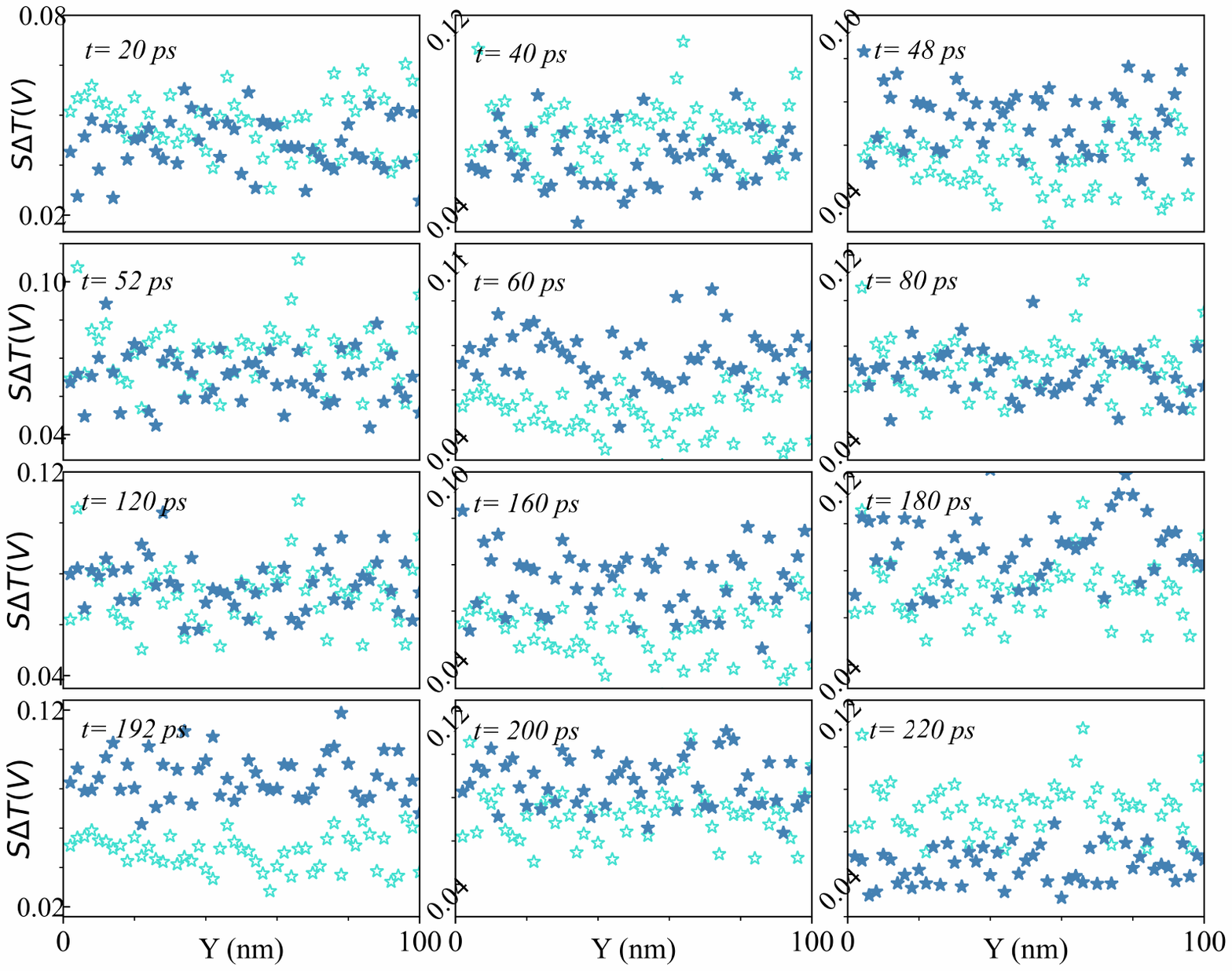}
		\caption{\label{vol2} The S$\Delta$T behavior of SnS$_2$ and SnSSe monolayers relative to the Y-value at different times.}
	\end{figure}

	At last step, considering all the preceding findings, the parameters involved in reaching higher energy conversion from the thermal energy directly into the electrical energy are examined. The generated Seebeck voltage for two materials of SnSSe and SnS$_2$ under the influence of the source heat of 1$\times$10$^{12}$ W/m$^{3}$, is plotted in Fig. \ref{vol2}. As said before, the Seebeck voltage is found through the relation $\Delta$V=S$\Delta$T. It is shown in Fig. \ref{Tprofile-heating}, during the heating at first 200 ps, the hot region is located at the left side where the heat is implied, and the cold zone is at the right side. The results confirm that throughout the first 200 ps, however the obtained efficiency for the SnSSe monolayer is almost one-half times larger than the one for SnS$_2$, the generated voltage, dependent also on the Seebeck coefficient, for two materials, is nearly the same, even at some times, slightly larger for less efficient predicted SnS$_2$. This finding confirms the inefficient definition of the maximum thermoelectric efficiency of the materials. In addition, the Fig. \ref{vol2} side by side with the Figs. \ref{kl-zt} and \ref{efficiency}, establishes that neither high zT nor large $\eta$ is not a guarantor for the obtained voltage through the device power conversion. That is to say, the phenomenon affirms that the higher value Seebeck coefficient, which can be assigned to the charge-carrier diffusion and the phonon drag, in cooperation with the larger temperature gradient, and also higher reached maximum temperature in a hot region due to the influence of the applied wasted heat, are responsible for larger value voltage and accordingly the better thermoelectric situation. 
	
	\section{Conclusions}
	\label{Sec.6}
	
	In this study, the heat transport in two materials of Janus SnSSe and SnS$_2$, which are candidates for efficient thermoelectric utilization, has been numerically investigated within the non-equilibrium Monte Carlo simulation of the phonon Boltzmann equation. In the first place, the phonon analysis has been performed. It is obtained unlike the suitable replacements for silicon channel in MOSFET designs, where the TA phonons play a unique role in thermal transport, in thermoelectrically efficient materials the contribution of TA phonons is much less. Consequently, the lack of the almost fast and high frequency transverse acoustic mode phonons for efficient thermoelectric materials causes them to reach the very high maximum temperature in response to the applied waste heat. This is the case for both two materials of SnSSe and SnS$_2$ monolayers where the TA phonons have the least contribution in heat propagation. Additionally, for two thermoelectrically efficient materials, the temperature profile in the XY plane has been investigated during the heating and cooling time. Immediately after applying the wasted heat, the hotspots are formed. The temperature gradient has been thoroughly investigated in Janus SnSSe and SnS$_2$ monolayers. At the beginning, the heat propagation through the SnSSe is slightly slower than that of the SnS$_2$ which is attributed to the lower thermal conductivity. In other words, as much as the heat propagation is gradual, the gradient between the hot and cold zones becomes larger. The case differs when the time passes and the thermal conductivity decreases with time and consequently the temperature. So, with time growing, the temperature gradient inside the SnS$_2$ becomes also somewhat more significant. The temperature difference is directly related to the Seebeck voltage, so the material with a higher discrepancy becomes more suitable for thermoelectric applications. Also, the thermoelectric efficiency, alongside the zT and K$_L$, and the value of the obtained Seebeck voltage is calculated. It is found, the much higher zT and $\eta$ do not result in a larger generated voltage. In brief, it is ascertained that the inherent Seebeck coefficient, and the temperature gradient, together with the obtained maximum temperature, are introduced to be responsible for the higher thermoelectric energy conversion.

\end{document}